\documentstyle[prl,aps,epsf,multicol]{revtex}
\begin{document}
\newcommand{\goodgap}{%
 \hspace{\subfigtopskip}%
 \hspace{\subfigbottomskip}}
\newcommand{\sia}{$\partial\sigma/\partial a$}
\newcommand{\sic}{$\partial\sigma/\partial c$}
\newcommand{\siu}{$\partial\sigma/\partial u$}
\newcommand{\kvec}{{\bf k}}
\newcommand{\vv}{\!\!\!\!\!\!}
\newcommand{\vareps}{\varepsilon}  
\newcommand{\vare}{\varepsilon}  
\newcommand{\eps}{\epsilon}  
\newcommand{\De}{$\Delta$}
\newcommand{\de}{$\delta$}
\newcommand{\mc}{\multicolumn}
\newcommand{\be}{\begin{eqnarray}}
\newcommand{\ee}{\end{eqnarray}}
\newcommand{\Pvec}{{\bf P}}

\draft 
\title{Roughening of close-packed singular surfaces}  
\author{Federica Trudu,$^{1}$\cite{p_add1} Vincenzo 
Fiorentini,$^{1,2}$  Paolo Ruggerone,$^{1}$
and Uwe Hansen$^{2}$\cite{p_add2}}

\address{(1) Istituto Nazionale per la Fisica della Materia and Dipartimento 
di Fisica, Universit\`a di  
Cagliari, Italy\\
(2) Walter Schottky Institut, TU  M\"unchen, Garching, Germany}
\date{Oct 19, 2000}
\maketitle
\begin{abstract}
An upper bound to the roughening temperature of a close-packed  singular 
surface, fcc Al (111), is obtained via free energy calculations  based on  
 thermodynamic integration using the  
embedded-atom interaction model. Roughening of Al (111) is predicted to
occur at around 890 K, well below bulk melting (933 K), 
and it should therefore be observable, save for possible kinetic hindering.
\end{abstract}

\begin{multicols}{2}
Roughening \cite{frank,hooge0,hooge1} is one of the most fundamental phase
transitions at  surfaces, yet probably the most elusive.
The roughening of   vicinal surfaces \cite{hooge0,hooge1}  is
generally accepted to be  a transition of infinite order
of the Kosterlitz-Thouless \cite{K-S} class.
The extremely weak free-energy divergence at 
the critical point implies
frustratingly slow variations in space and time of
whatever order parameter is chosen to characterize the transition.
This makes predictions on roughening a challenge for atomistic
simulations techniques, this being not the last of  the reasons 
why statistical mechanics models \cite{models} have traditionally
been the dominant approach to this problem.

The roughening of singular faces poses additional problems.
 Vicinal surfaces roughen as the 
(mostly configurational) entropic free energy related to step
meandering prevails  over the cost of  step and kink formation; 
on vicinals, where steps already exist by construction, this
occurs generally  at temperatures well below melting. 
Singular-face roughening, on the other hand,
 requires  step formation  to begin with.
Singular faces, therefore, roughen at much higher temperatures, so 
much so that roughening is thought to be preempted by melting in most
cases, especially on close-packed faces.

Here we use a simple approach to predict the roughening transition
temperature  of a singular  
surface, based on free energy calculations
performed with an atomic-level finite-temperature
 simulation technique  (the embedded atom method coupled with Monte Carlo 
thermodynamic integration). 
We calculate the free energies of several vicinals 
to the singular face, and estimate the temperatures at which the
  free energy of each vicinal becomes lower than the singular.
Since roughening is  phenomenologically identified with the appearance
of hills and valleys of arbitrary height on the surface, we assume
that roughening will be fully developed at the temperature at which  the
steepest and most costly vicinal will be favored over the 
low-index face. To obtain an internally consistent and low-error-bar
 estimate, we calculate the crossing temperatures
of the free energies of  several vicinals with progressively shorter terraces, 
with the free energy  of the singular surface; we then
 obtain T$_{\rm R}$ as  the extrapolated crossing temperature
of the shortest/most costly vicinal. To be definite,  here we  estimate 
an  upper bound  to T$_{\rm R}$ for any Al surface, and find it to
be $\sim$ 890 K, well below the bulk melting temperature of 933 K.

 To obtain such upper bound,  we study Al (111),  which
  is expected to have the   highest roughening
temperature among the low-index faces, being the most
closely packed. Also, it is
 stable \cite{denier}  up to the bulk melting temperature, 
and   predicted  to sustain overheating \cite{tmelt}.
 The vicinals of Al (111) we consider here
are Al (8\, 8\, 10), Al (557) and Al (335), obtained by miscut of
the (111) plane at an angle of $\sim$ 1, 9, and 14 degrees
respectively.  There exist two kind of steps on Al (111), namely the 
{111}-facetted and the {100}-facetted. The latter are energetically
more costly, and  our vicinals belong to this second class. In
the notation of Lang\cite{lang},  bearing  out 
directly the interstep distance, these faces are denoted as
[9(111)$\times$(100)], [6(111)$\times$(100)] and [4(111)$\times$(100)],
respectively, meaning (say) 6 rows of a (111) face separated by a
(100)-faceted step. These vicinals  lay on the (111)-(100) line of the 
stereographic map of the fcc lattice \cite{camp}. The steepest vicinal
 on this line is  Al (113), or [2(111)$\times$(100)] : its
appearance should set the occurrence of fully developed roughening.
Here we first simulate straight-step vicinals, and then estimate the
correction due to kink formation by simulating one kinked vicinal. 

Free energies are calculated via the embedded atom method and
thermodynamic integration. The embedded atom method \cite{DB} is
a fairly reliable method to predict structural and thermal properties
of metals. Its main 
advantage is its moderate  computational cost, and ensuing  high
numerical accuracy achievable within the method's bounds. The
disadvantages are essentially that
the choice of materials to be simulated is restricted by the
availability of accurate potentials (constructing  which
 is a science in itself), and that 
the embedded atom method, by its nature of effective  interatomic
potential, is not as accurate  
as first principles methods. This inherent inaccuracy is attenuated
for Al by the highly refined parameterization of Ercolessi and 
Adams\cite{Erc}, built to reproduce a large database of ab initio
energy and force calculations. Recently \cite{fiore} the
Ercolessi-Adams model has been further refined to cure 
minor inaccuracies in the description of surface diffusion and
high-energy scattering.  

Thermodynamic integration is adopted because the roughening transition
occurs (if at all) well above the Debye temperature  ($\sim$ 400 K for
Al bulk), and it is therefore imperative to properly include  anharmonic
effects in the free energy  of the relevant surfaces. While 
useful at lower temperatures, the commonly  adopted quasi-harmonic
approximation is not very reliable at  high temperature, as shown by
recent simulations\cite{fiore} on Al (100). 
In thermodynamic integration
\cite{TI}, the potential energy of the system is progressively
switched on, through a parameter $\lambda$, starting from a reference
system whose free energy is known:
\be\label{H(l)}  
\rm
 V \, (\lambda) =\lambda\, W - (1-\lambda)\,  U_h
\ee
with W and U$_{h}$ the potentials of the actual system and of an
harmonic crystal. Since \cite{TI},
\be\label{df}
\rm
\frac{\partial F}{\partial \lambda}= \langle U_h - W \rangle _{\lambda }, 
\ee
 the free energy at a fixed temperature T$_{\rm ref}$ is
\be\label{F}
\rm
{\rm F}_{\rm ref} = \rm F_{\lambda=1}=F_{\lambda=0}+\int_{0}^{1} \langle U_{h}-W\rangle _{\lambda
}\,d\lambda \, .
\ee
 The integrand is calculated  by Metropolis canonical Monte Carlo, 
and the $\lambda$=0 value is known by  
construction. By the  thermodynamical
free energy-enthalpy relation
\be\label{df/dT}
\rm
\frac{d}{d\, T}\left(\frac{F}{T}\right)=-\frac{H}{\,\ T^2}
\ee
the free energy in the interval  [T$_{\rm ref}$,T] is
\be
\rm
F (T)=T\, \left[ \frac{F_{{\rm ref}}}{T_{{\rm ref}}}-\int_{T_{{\rm
ref}}}^{T}\!\frac{H}{\,\ T^2}dT\right].
\ee
The integrand is calculated again by canonical Monte Carlo. The surface
free energy per unit area is 
\be
\rm
F_{\rm surf}\, (T)=\frac{1}{2\, A}\, \left[
F_{\rm slab}\, (T) - N\, F_{\rm
bulk}\, (T)
\right]
\ee
where
 F$_{\rm bulk}$ (T) is the bulk free energy per atom,  F$_{\rm
slab}$ (T) is the free energy  of the N-atom  simulation slab,
and A its surface area. Whenever appropriate,
thermal expansion  is accounted for with an expansion coefficient
given by the ratio a(T)/a(0) of the lattice constants at
temperatures T and zero, obtained   by  an NPT simulations \cite{fiore}.
Free energies are calculated in supercells containing 
450 to 600 atoms depending on the orientation. Each 
Metropolis Monte Carlo run
 was $\sim$3$\times$10$^7$ steps long. We estimate error bars
of  0.5\% in the surface free energy, and
 about $\pm$10 K in the
crossing points and T$_{\rm R}$. 

\narrowtext
\begin{figure}
\epsfclipon
\epsfxsize=8.5cm
\epsffile{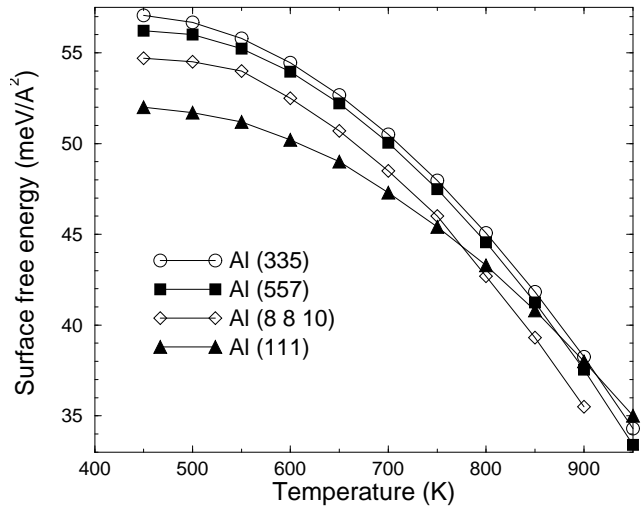}
\caption{Free energy temperature dependence for Al (111) and vicinals.}
\label{fig.deltaf}
\end{figure}

In Fig. \ref{fig.deltaf} we report the
free energy vs temperature for Al (111) and vicinals.
 At first, Al (111) is favored.
At higher temperatures, vicinals with progressively shorther
 terraces become  favored free energy-wise.
The crossing points are T=773 K for  Al (8\,8\,10),
 T=875 K for  Al (557), and T=914 K for Al (335). 

\begin{figure}
\epsfclipon
\epsfxsize=8.5cm
\epsffile{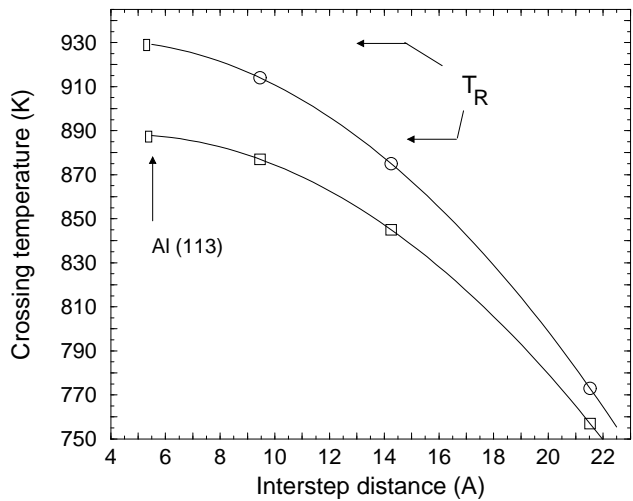}
\caption{Crossing points of surface free energies of
vicinals and singular surface. Upper curve: straight steps;
lower curve: kinked steps. T$_{\rm R}$ is
defined as the temperature corresponding to
the Al (113) interstep distance.}
\label{fig.extrapol}
\end{figure}

The  crossing points in Fig.\ref{fig.deltaf}
 tend to ``accumulate'' towards a finite value as the
terraces become progressively shorter. This suggests to identify 
T$_{\rm R}$ with the ``accumulation point'' of this sequence. 
To quantify it, we fit a polynomial through
the crossing points just obtained as function of interstep  distance,
and define T$_{\rm R}$  as the temperature value corresponding to the
interstep distance on the vicinal surface with the shortest terrace
within  our class  of  (100)-faceted, straight-stepped vicinals,
namely  Al (113), whereby the interstep 
distance is $\sim$ 5.5 \AA. At  T=T$_{\rm R}$ as just defined, 
all the vicinals (within our restricted class)
 are favored over  Al (111), so that arbitrarily large and composite
fluctuation can  appear in the surface profile. The result is
displayed by the upper curve  in 
 Fig. \ref{fig.extrapol}: the roughening temperature
estimate  is  930 K, very close to the 
 melting temperature of Al bulk (theoretical \cite{tmelt}: 939 K;
 experimental \cite{denier}: 933 K).

To refine the prediction, we first note that only undefected
straight steps have been  considered so far. On the other hand, 
 at finite temperature kinks will form on steps. Kinks affect 
the free energy of the stepped surface both
 indirectly because their very existence allows step meandering,
and  directly via their  formation internal  energy, and  vibrational
entropy due to their  vibrational modes. We consider that the latter
free-energy variation will be captured accurately by a simulation.
We neglect instead the step meandering-related configurational entropy,
based on  previous work on vicinal surfaces \cite{fr-st} suggesting
 that the configurational contribution is negligible compared to
the vibrational below the roughening transition.

\begin{figure}
\epsfclipon
\epsfxsize=8.5cm
\epsffile{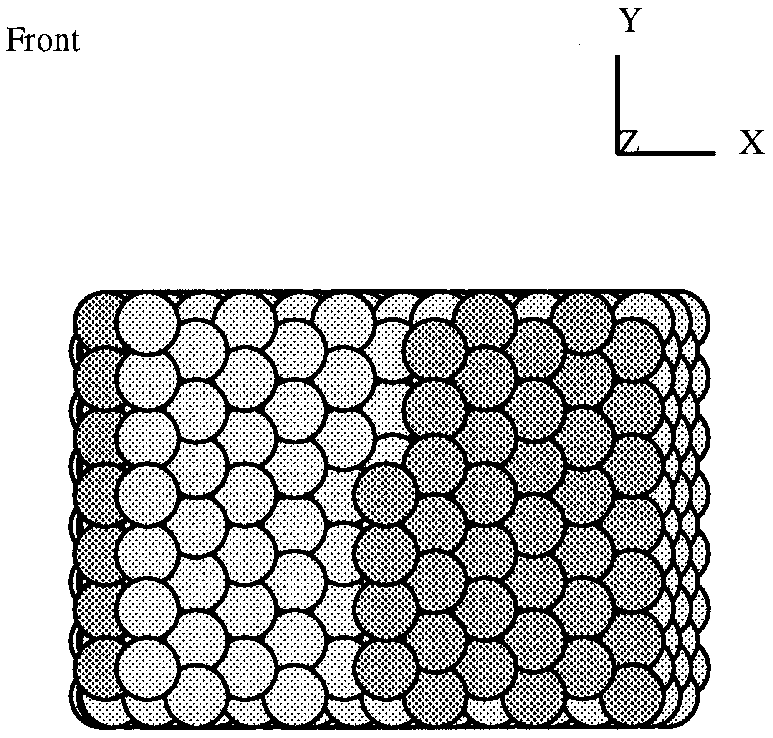}
\caption{Top view of kinked Al (557) as
studied in free energy calculation. 
As all other cells, it contains two 
periodically-repeated steps
per side.}
\label{fig.kink}
\end{figure}

\begin{figure}
\epsfclipon
\epsfxsize=8.5cm
\epsffile{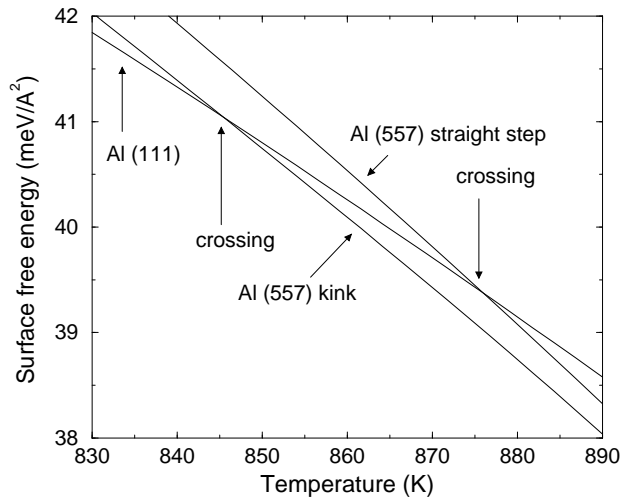}
\caption{Lowering of the free energy
 crossing point of Al (557) with Al (111) due to the presence of
kinks.} 
\label{fig.cross}
\end{figure}

The simulation of vicinal 
surfaces with kinks is demanding in periodic boundary conditions;
here we restrict to a single case, kinked Al (557),
chosen because of its favorable geometry. Each side of the
simulation slab,  depicted in Fig. \ref{fig.kink}, contains one straight 
and one kinked step. The latter exhibits two kinks,
with a relatively low linear density  of 
0.05 \AA$^{-1}$. The number of atoms is preserved by this
procedure, as required by numerical considerations.  As shown in Fig.
\ref{fig.cross}, the kinked 
 Al (557)  turns out to have a  crossing point with Al (111)  at T=845
K, with a  reduction of 4\% over the straight-step value.
Assuming that  the other crossing points
are lowered by about the same amount due to kinks, 
and applying the same procedure as before, we 
find  T$_{\rm R}$= 887 K (lower-laying 
curve in Fig. \ref{fig.extrapol}).  This is a strong  upper 
bound because accounting for lower-cost (111)-faceted steps   should
lower this figure. In addition, account for meandering
will also lower (moderately) our estimate.

Roughening  has not been reported for any (111)  face so far. 
The  predicted T$_{\rm R}$ is rather close to, but lower than 
the melting temperature, so it is quite  conceivable that roughening 
of Al (111) could be observed. Our prediction concerns 
energetics, however. Kinetic effects are not considered in any way.
However,  Al (111)  was observed \cite{denier}  in Medium
Energy Ion Scattering experiments to remain stable up to the
melting temperature. Also, molecular dynamics simulations \cite{tmelt}
 showed Al (111) to be stable for at least 2 ns up to 1088 K,
or 150 K above bulk melting. While the length and time scales
accessible in  simulation are not comparable with those of relevance
in roughening, this is an indication that   kinetics may play a role,
slowing down or hindering the transformation. Thus, it is possible
that  experiments aiming at the observation of  the roughening of Al
(111) predicted here, may have to observe the surface over time spans of
hours, or produce ``nucleation'' defects by e.g. nanoindentation. 
 
As a further check on  the predictions based on the
embedded atom Al potential, reinforcing the plausibility of our estimate, 
we calculate  T$_{\rm R}$ for vicinals within the Terrace-Ledge-Kink (TLK) 
model of
Villain {\it et   al.} \cite{vgl}, through the relation
\be
\rm
K =  \frac{W_m}{\rm k_{\rm B} T_{\rm R}} 
e^{W_0/\rm k_{\rm B} T_{\rm R}} .
\ee
Here 
W$_{\rm m}$ is the energy needed to move a step by one 
row towards  a neighboring step
 m+1 atomic rows away, and W$_0$ 
is the kink formation energy. This expression is valid for
 W$_{\rm m} < {\rm T} < {\rm W}_0$, which is  the case here. The
value of $K$ depends upon the details of the underlying  theory, and
it equals 2 for the original TLK model; values of 2  for Cu (113)
\cite{s-s} and 2.1 for  Ag (115) \cite{hooge1}  have been suggested
based on experiments or MonteCarlo simulations on vicinals. We 
evaluate these parameters from total energy calculations
on slabs containing at least 5 steps per slab side,
and comprising from 1700 to 4000 atoms depending on orientation. 
The parameter W$_{\rm m}$ is calculated removing one complete atomic 
row of step-edge atoms. If N is the total number of atoms, 
and L that of step-edge atoms, the total energy for row
remowal is 
\be  
\rm L\, W_{m}=E_{N-L}-[E_{N}-LE_{b}],
\ee
with E$_{\rm N-L}$ and E$_{\rm N}$  the internal energy of system
after and before row  removal, and E$_{\rm b}$  the bulk energy
per atom.  W$_{\rm m}$ is thus defined per atom. For a kink we only remove
half a row, creating two kinks:  
\be
\rm
2\, W_{0}= E_{N-L/2} - [E_{N}-(L/2)\, E_{b}] - W_m.
\ee
with $\rm E_{N-L/2}$ the internal energy of the slab after half-row
removal.

For Al (335) we find W$_{\rm m}$=3 meV, W$_{0}$=112 meV, T$_{\rm R}$= 411 K;
for Al (557) we find W$_{\rm m}$=1 meV, W$_{0}$=108 meV,  T$_{\rm R}$= 314 K;
 for Al (8\, 8\, 10) we find W$_{\rm m}$=0.1 meV, $W_{0}$=106 meV, and
 T$_{\rm R}$= 209 K. These values are quite comparable with results of
previous investigations on stepped metal surfaces\cite{G-H,ham-lyn}.  
Our numbers for Al (335) are  compatible with those inferred from STM
measurements on Ag (115)\cite{hooge2}, which has the same step-step
separation: W$_{\rm m}$= 3 meV, W$_{0}$=114 meV, and T$_{\rm R}$=427 K.
[The (115) face
consists of (111)-faceted steps separated by a (100) terrace four atomic
rows wide, whereas the (335, has (100) steps and (111) terraces.] 
Concerning T$_{\rm R}$  of Al (111), den Nijs {\it et al.} 
observed\cite{conr} roughening of Ni (115) at about 200 K, and estimated 
420 K for the roughening of Ni (100), the nearest singular face on
the stereographic plot.
Our value of 412 K for Al (335) similarly suggests that our upper bound 
of 890 K for the associated singular (111) is quite plausible.
Our predictions for both singular and vicinal faces await experimental
verification. 

In summary we have calculated an upper bound to the roughening
temperature of a singular metal surface using an atomistic simulation
method. Our results for Al (111) suggest that roughening may occur
appreciably below melting, and  be therefore  observable,
save for kineting hindering.

VF thanks the Alexander von Humboldt-Stiftung for supporting 
his stays at the Walter Schottky Institut.

\end{multicols}
\end{document}